\documentclass[structabstract]{aa}  
\usepackage{graphicx}
\usepackage{txfonts}
\usepackage{natbib}
\bibpunct{(}{)}{;}{a}{}{,} 

\usepackage{lscape}

\usepackage{longtable}
\begin{document}
   \title{The bright galaxy population of five medium redshift clusters}
   \subtitle{II. Quantitative Galaxy Morphology}

   \author{B. Ascaso  \inst{1,3},  J. A. L. Aguerri \inst{2} , M. Moles \inst{3}, R. S\'anchez-Janssen 
          \inst{2} and D. Bettoni \inst{4}  
          }
   \institute{Department of Physics, University of California, Davis, 
   	    One Shields Avenue, Davis, CA 95616, USA\\
              \email{ascaso@physics.ucdavis.edu}
         \and
             Instituto de Astrof\'isica de Canarias, C/ Via
	  Lactea S/N C.P: La Laguna, Spain\\
             \email{jalfonso@iac.es, ruben@iac.es}
            \and
            Instituto de Astrof\'isica de Andalucia, Camino
         Bajo de Huetor 50, C.P: 18008 Granada, Spain\\
          \email{moles@iaa.es}
          \and
          INAF- Osservatorio Astronomico di Padova, Vicolo 
          Osservatorio 5, 35122, Padova, Italy\\
             \email{daniela.bettoni@oapd.inaf.it}
             }

   \date{Received 29 December 2008 / Accepted 2 July 2009}

 
  \abstract
   {}
   {Following the study already presented in our previous paper, based on the Nordic Optical Telescope (NOT) sample, which consists of five clusters of galaxies within the redshift range 0.18 $\leq$ z $\leq$ 0.25, imaged in the central 0.5-2 Mpc in very good seeing conditions, we have studied the quantitative morphology of their bright galaxy population}
   {We have analyzed the surface brightness profiles of the galaxy population in those clusters. Previously, we have performed simulations in order to check the reliability of the fits. We have also derived a quantitative morphological classification.}
   {The structural parameters derived from these analysis have been analyzed. We have obtained that the structural parameters of E/S0 galaxies are similar to those showed by galaxies in low redshift clusters. However, the disc scales are different. In particular, the scales of the discs of galaxies at medium redshift clusters are statistically different than those located in similar galaxies in the Coma cluster. But, the scales of the discs of galaxies in medium redshift clusters are similar to nearby field galaxies.}
   {The results suggest that the evolution of the disc component of galaxies in clusters is faster than in field ones. Mechanisms like galaxy harassment showing timescales of $\sim 1$Gyr could be the responsible of this disc scale evolution. This indicates that spiral galaxies in clusters have suffered a strong evolution in the last 2.5 Gyr or that Coma is in some way anomalous.}

 \titlerunning{Bright galaxy population of five medium redshift clusters. II}
 \authorrunning{Ascaso et al.}
   \maketitle
%

\section{Introduction}


The brightest galaxies in the central part of clusters of galaxies have been an object of study for years \citep{kormendy77,dressler80,merritt84,caon93,jorgensen94,kauffmann94,bower98,fasano00,aguerri04,deLucia07,ascaso08}. Most of these studies have investigated their link with the formation of their host halo leading to two main scenarios: the monolithic \citep{merritt84,bower98} and the hierarchical \citep{kauffmann94,deLucia07}. The former assumes the cluster to be formed first and, consequently, the galaxies are not suffering transformations after the cluster collapse. The latter, on the other hand, implies that the galaxies were formed earlier than the cluster and therefore, environmental effects or interacting mechanisms such as harassment \citep{moore96}, gas-stripping \citep{gunn72,quilis00}, starvation \citep{bekki02}, or merging \citep{aguerri01,eliche-moral06} are able to alter the galaxy population.

The galaxy population has a bimodal nature based on their stellar population and the shape of their surface brightness profiles. As \cite{driver06} showed, the redder and more compact objects are usually early-type systems, while the late-type galaxies are, generally, bluer and less concentrated profiles. Recent results, \citep{ascaso08b,bell08,skibba08,cameron09} have pointed out that this bi-modal behaviour for field and cluster galaxies can be translated in the color-log($n$) plane, where $n$ is the S\'ersic index \citep{sersic68}, up to redshift 1. This result agrees with the hierarchical clustering scenario as the late type galaxy population would be formed by cooling the gas in the dark matter halos and, as a consequence, the early type population would be formed by merging of late type galaxy members or by dry-merging of the early type galaxies.

Many works related to the evolution of the galaxy population in clusters of galaxies have also shown a decrease of the S0/E fraction with redshift. This decrement is mainly due to the variation with redshift of the S0 fraction, while the Elliptical fraction remains constant up to z $\sim$ 0.8, \citep{fasano00,dressler97,postman05,desai07}. This fact indicates a different time scale for the process of formation of elliptical and lenticular galaxies.  However, \cite{jorgensen94} studied the nature of elliptical and S0 galaxies in the Coma cluster, suggesting that they are a continuum class of objects as a distribution of bulge-to-disk ratios. Subsequent studies have given support to this fact \citep{jorgensen96,krajnovic08}. 

Likewise, a number of  papers reflect a continuity in the parameter space for bright elliptical and dwarf galaxies in clusters, \citep{sandage78,graham03,gutierrez04,aguerri04,ascaso08} suggesting that these objects could be a continous family. However, recent studies have been devoted to the study of the physical differences between elliptical  and spheroidal or dwarf elliptical galaxies \citep{aguerri05,aguerri09,kormendy08} suggesting recent evolution in their structural parameters. For example, \cite{kormendy08} have shown that the bright ($M_{VT} \le -21.66$) elliptical galaxies in the Virgo cluster have cuspy cores, rotate slowly, have anisotropic velocity distributions, boxy isophotes and S\'ersic values, $n$, larger than 4, while the faint ellipticals ($-21.54 \le M_{VT} \le -15.53$) do not have cores, rotate much faster, usually have more isotropic velocity distributions, disky isophotes and smaller S\'ersic parameters.  In addition, they confirm that the biggest elliptical galaxies have X-ray emitting gas whereas the smallest ones have a lack of it.  Therefore, the existence of larger samples of faint dwarf galaxies at medium-higher redshift  is necessary to study their evolution.

The inspection of the galaxy surface brightness and  main structural parameters has been studied in several works for different samples. Those samples, have been restricted to local clusters \citep{caon93,jorgensen94,aguerri04,gutierrez04} or to galaxy samples preselected by morphological type, (e.g. early types \citep{caon93,jorgensen94,graham03} or late types \citep{dejong96,graham01,graham03,mollenhoff04}).  In addition, larger samples of field galaxies have been analyzed. For example, \cite{trujillo04} presented  quantitative structural parameters in the V-band rest-frame for all galaxies with $z < 1$ and $I814(AB) < 24.5$ mag in the Hubble Deep Fields North and South. Nevertheless,  the present number of works devoted to the study of global samples of galaxies in clusters at medium redshift range is small, \citep{fasano00,trujillo01,balogh02}. In this work, we have studied the surface brightness of the whole galaxy population in a sample of clusters of galaxies at medium redshift.

Our data consists of five clusters observed with the Nordic Optical Telescope (NOT) in very good seeing conditions with two filters in a range of redshift from 0.18 to 0.25, where there are very few cases of analysed clusters due to difficulties in the deepness and quality of the observations. The analysis of the properties of such clusters can provide a new perspective in the evolutionary trends of clusters at that range of redshift, as well as, the extension of the properties of other samples at low redshift to high redshift (HST).

In this paper, we have continued the analysis of the properties of the brightest galaxies ($m_r \le 20$) in the central part of  this cluster sample at medium redshift. We have analyzed their surface brightness, and performed an study of their structural parameters. We have also extracted their quantitative morphology and compared this to their visual morphology already derived in \cite{ascaso08}.

The structure of this paper is as follows. In section 2, we describe our sample and its analysis and we explain how we have fitted the surface brightness analysis into a one- S\'ersic profile \citep{sersic68} or a two-component profile, S\'ersic+ Exponential profile, \citep{freeman70}. In section 3, we explain and show the results of the simulations for the establishment of the range of the parameters where we can fit the different models. In section 4, we describe the galaxy classification and analyze the structural parameters extracted from the surface brightness fits for bulge and disc galaxies. Finally, we show the discussion of the results and conclusion in section 5. Throughout that paper we have adopted a $\Lambda$CDM cosmology: H$_0$=71 km s$^{-1}$ Mpc$^{-1}$, $\Omega_m$=0.27 and $\Omega_{\Lambda}$=0.73.


\section{Data analysis}

We have extended our study to the data previously presented in \cite{ascaso08} of five clusters of galaxies imaged with  the Stand Camera of the 2.5m Nordic Optical Telescope (NOT) located at the Roque de Los Muchachos Observatory (La Palma). The main characteristics of the sample are collected in Table \ref{tab:all}.

We refer the reader to \cite{fasano00} for more extended details about the observations and the data reduction process. The detection and extraction of the galaxy sample is widely explained in \cite{ascaso08}.

 \subsection{Two dimensional surface brightness fit}

The surface brightness of the galaxies in our medium redshift clusters were modelled using one or two photometrical components. The fits were carried out using the automatic fitting routine (GASP-2D) developed and successfully validated by \cite{jairo08}. The surface brightness of those galaxies modelled with only one component was described by a S\'ersic law \citep{sersic68}, while the surface brightness of those galaxies fitted with two photometrical components were described by a S\'ersic law plus an exponential one \citep{freeman70}. 

The fits of the galaxies were fully two-dimensional. The photometrical galaxy components were characterized by elliptical and concentric isophotes with constant (but possibly different) ellipticity and position angle. We have asumed a cartesian coordinates system $(x,y,z)$ with origin in the galaxy center, the x-axis parallel to the direction of the right ascension and pointing westward, the y-axis parallel to the direction of declination and pointing northward, and the z-axis along the line-of-sight and pointing toward the observer. The plane of the sky is confined to the $(x,y)$ plane, and the galaxy center is located at the position (x$_{o}$, y$_{o}$).

The S\'ersic law has been extensively used in the literature to model the surface brightness of elliptical galaxies \citep{grahamg03,graham05}, bulges of early and late-type galaxies, \citep{andredakis95,prieto01,aguerri04,mollenhoff04}, the low surface brightness host of blue compact galaxies \citep{caon05,amorin07,amorin09}, and dwarf elliptical galaxies, \citep{binggeli98,grahamg03,aguerri05}. The radial variation of the intensity of this law is given by:

\begin{equation}
I(r)=I_e10^{-b_n[(r/r_e)^{1/n}-1]}
\end{equation}

where $r_{e}$, $I_{e}$, and $n$ are the effective radius, the intensity at $r_{e}$ and a shape parameter, respectively. The value of $b_{n}=0.868n-0.142$ is coupled to $n$ so that half of the total luminosity is within $r_{e}$, (see Caon et al. 1993). The isophotes of the S\'ersic models are concentred ellipses centred at $(x_{o}, y_{o})$ with constant position angle $PA_{b}$ and constant ellipticity $\epsilon_{b}=1-q_{b}$. Thus, the radius $r_{b}$ is given by: 

\begin{eqnarray}
r_{b}&=&[(-(x-x_{o}) sin PA_{b} + (y-y_{o}) cos PA_{b})^{2}  \nonumber \\
& & - ((x-x_{o}) cos PA_{b} + (y-y_o) sin PA_{b})^{2}/q_{b}^{2}]^{1/2}  
\end{eqnarray}

Hereafter, we will call 'bulge'  the photometric galaxy component fitted by a S\'ersic law in those galaxies fitted with two components.

On the other hand, the exponential law has been used in the literature to model the surface brightness profile of the discs showed by spiral galaxies (e.g. \cite{aguerri05} and references therein). This law was proposed by \cite{freeman70} and is given by:

\begin{equation}
I(r)=I_0 e^{-r_d/h}
\end{equation}

where $I_{0}$ and $h$ are the central intensity and scale length, respectively. Similar to the photometrical component modelled by a S\'ersic law, we have considered that the disc isophotes are ellipses centered on the galaxy center ($x_{o}, y_{o}$) with constant position angle $PA_{d}$ and constant ellipticity $\epsilon_{d}=1-q_{d}$, given by the galaxy inclination $i=arcos(q_{d})$. Thus, the radius $r_{d}$ is given by:

\begin{eqnarray}
r_{d}&=&[(-(x-x_{o}) sin PA_{d} + (y-y_{o}) cos PA_{d})^{2} \nonumber \\
& & - ((x-x_{o}) cos PA_{d} + (y-y_{o}) sin PA_{d})^{2}/q_{d}^{2}]^{1/2}  
\end{eqnarray}

The GASP-2D routine fits all free parameters iteratively using a non-linear least-squares minimization method. It was based on the robust Levenberg-Marquardt method \citep{press92}. During each iteration of the fitted algorithm, the seeing effect was taken into account by convolving the model image with a circular point spread function (PSF) extracted from the images (see \cite{jairo08} for more details about the fitting routine).

As many authors have recently explored, \citep{gadotti08,cameron09}, the presence of bars in the galaxies may result in poor bulge fits if not correctly modelled. Nevertheless, we have taken care of that by examining carefully the residuals.

\section{Simulations}

One of the advantages of the quantitative morphology is that the accuracy of the obtained results can be tested by simulating artificial galaxies similar to the real ones. We have created a large number of artificial galaxies with one and two galactic components described by the previous equations. These modeled galaxies are similar to the galaxies observed in our medium redshift galaxy clusters.

We generated 5000 images of galaxies with a S\'ersic component. The total magnitude, effective radius, shape S\'ersic parameter, and ellipticity of the simulated galaxies were similar to those from real galaxies. They were asigned randomly to the models, and their values were in the ranges:

\begin{eqnarray}
18\leq m_{r} \leq 21  ; \quad 0.5 kpc \leq r_{e} \leq 4 kpc  ; \quad  0.5\leq n \leq 6  ;  \quad 0.3 \leq q_{b} \leq 1
\end{eqnarray}

We have also generated 5000 galaxies with two photometric components: S\'ersic and exponential. These artificial galaxies have a central photometric 'bulge' component, modeled by a Sersic law, and an external 'disc' component, modeled by an exponential law. The total magnitude of these galaxies span a range $18\leq m_{r} \leq 21$. The contribution to the total light of the galaxies by the bulge and disc components is given by the bulge-to-total light ratio. This parameter spreads over  the range $0\leq B/T \leq 1$. The bulge parameters of the simulated galaxies were:

\begin{equation}
0.5 kpc \leq r_{e} \leq 4 kpc ; \quad  0.5\leq n \leq 6   ; \quad  0.2\leq q_{b} \leq 1
\end{equation}

The disc free parameters of the galaxies were:

\begin{equation}
1.75 kpc \leq h \leq 7 kpc  ; \quad 0.2\leq q_{d} \leq 1
\end{equation}

In order to mimic the same instrumental setup, we added a background level and photon noise to these artificial images similar to the observed images. They were also convolved simulating the seeing that we have in our observations. Finally, these simulated galaxies were fitted with identical conditions as the real ones. The simulated galaxies will be used for determining the errors of the fitted structural parameters.

\subsection{Galaxies with one photometrical component}

\begin{figure}[]
\centering
\includegraphics[clip,angle=90,width=1.25\hsize]{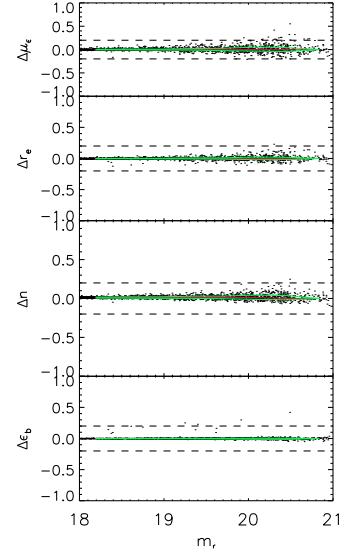}
\caption{Plots of the magnitudes versus absolute ($\mu_e$) and relative ($r_e$, $n$ and $\epsilon_b$) errors for the parameters of the S\'ersic  profile. The horizontal dashed lines are the 20\% of the error. The green  and red lines are the quartile and percentile of the error  respectively in bins.}
\label{fig:paramS}
\end{figure}

Figure \ref{fig:paramS} shows the errors of the free parameters of those simulated galaxies with only one component as a function of their magnitude absolute errors for $\mu_e$ and relative errors for $r_e$ ,$n$ and $\epsilon_b$). Notice that bright galaxies show small relative errors in the fitted parameters than faint ones. Consequently, the goodnes of the fit depends on the galaxy magnitudes or, due to the correlation between area and magnitude, on the galaxy area. The areas of the objects were computed as the number of pixels with higher signal than 1.5 times the $rms$ of the sky background of the images, and belonging to the intersection of the observed image and the fitted model. The last restriction was imposed in order to avoid wrong area measurements due to nearby objects. 

\begin{figure}[h]
\centering
\includegraphics[clip,angle=90,width=1.0\hsize]{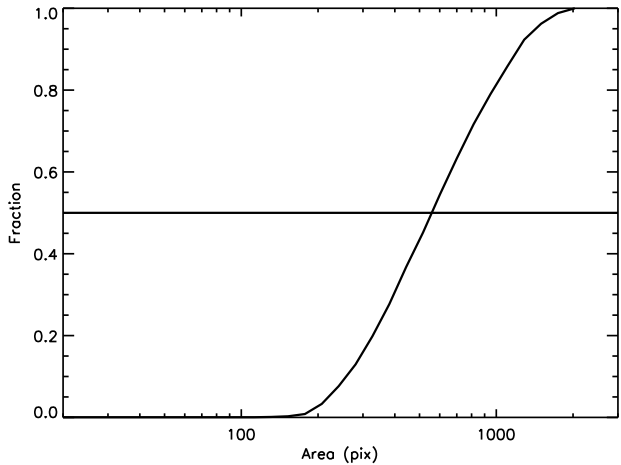}
\caption{Fraction of properly fitted galaxies versus area of the  one-component simulated galaxies.}
\label{fig:simu1}
\end{figure}

We have considered that a galaxy was properly fitted when all free parameters were recovered with relative errors less than 20$\%$. Figure \ref{fig:simu1} shows the fraction of properly fitted simulated galaxies as a function of the area. We defined the minimum area of properly fitted galaxies as the one at which the fraction showed in Fig. \ref{fig:simu1} is equal to 0.5. This correspond to 550 pixels for galaxies modeled with only one S\'ersic component. This means that more than 50$\%$ of the galaxies with areas larger than 550 pixels were properly fitted.

\subsection{Galaxies with two photometrical components}

\begin{landscape}
\begin{figure}[]
\centering
\includegraphics[clip,angle=90,width=1.0\hsize]{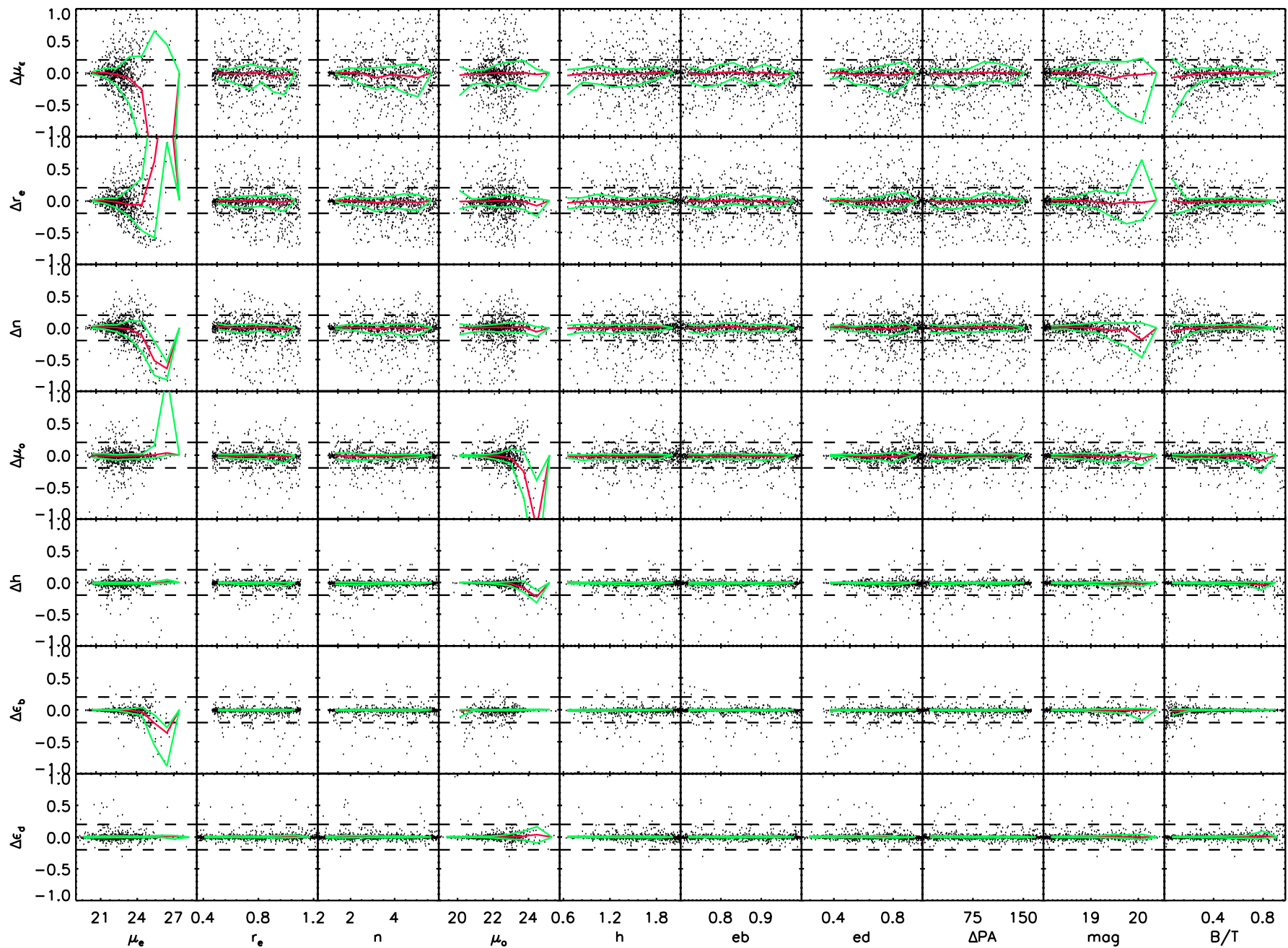}
\caption{Plots of the S\'ersic+ Disc profile parameters versus its  error for those galaxies with areas larger than 800 pixels (see text  for details). The lines are the same as  Fig. \ref{fig:paramS}.}
\label{fig:paramNR}
\end{figure}
\end{landscape}

Figure \ref{fig:paramNR} shows the relative errors of the fitted free parameters of the simulated galaxies with bulge and disc components except for $\mu_e$ and $\mu_0$, where we show the absolute errors. Notice that in general the disc parameters are better fitted than the bulge ones. It is also clear that those galaxies with large $B/T$ show larger errors in the disc parameters than in the bulge ones. In contrast, galaxies with smaller $B/T$ show larger errors in the bulge than in the disc. Indeed, the bulge and disc parameters  for faint galaxies ($m_{r}>20$) with low
surface brightness ($\mu_{0,D}>25.0$ or $\mu_{0,B}>25.3$ mag arcsec$^{-2}$), are not well fitted.

\begin{figure}[h]
\centering
\includegraphics[clip,angle=90,width=1.0\hsize]{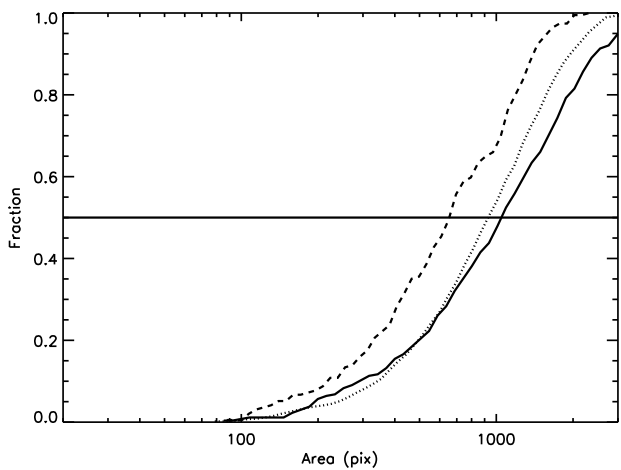}
\caption{Fraction of properly fitted bulge and disc simulated galaxies as a function of their area. The full line represents galaxies with $B/T<0.2$, the dotted line those with $0.2<B/T<0.8$, and the dashed line $B/T>0.8$.}
\label{fig:simu2}
\end{figure}

We have also considered that a galaxy was properly fitted if all free parameters were recovered with relative errors smaller than 20$\%$. Figure \ref{fig:simu2} shows the fraction of properly fitted bulge and disc simulated galaxies. In this case, the area at which at least 50$\%$ of the population is well fitted depends on their $B/T$ values. We have adopted in this case 800 pixels as the minimum area in order to obtain reliable fits.

\subsection{Number of components}

The simulations have showed that all galaxies with area larger than 800 pixels can be properly fitted. This area corresponds to galaxies brighter than $m_{r}=20$. Thus, all the galaxies down to $m_{r}=20$ were fitted with one and two components. As far as the galaxies with areas comprised between 550 and 800 pixels are concerned, we didn't include them in our analysis.

Usually, the $\chi^{2}$ value is used for deciding which is the best fitted model. Nevertheless, models with two components tend to have smaller values of $\chi^{2}$ than models with one component. This fact  is due to the different number of free parameters between both fits. Additionally, it is also possible that the model with the lowest value of  $\chi^{2}$ could not be a physical solution. What is more, luminous, well resolved galaxies with spiral arms or inner dust-lanes that can not be properly described by the model, tend to have larger values of $\chi^{2}$.  For these reasons, we have adopted an alternatively method in order to decide the best fitted photometric model. The method is similar to the used by \cite{allen06}. It is based on the  analysis of the surface brightness radial profiles of the fitted models. Our aim is that those galaxies finally fitted with two components should be 'classical' bulge and disc systems, i.e. their central regions should be dominated by the bulge component, while the disc dominates at large radial distances from the galaxy center. Galaxies with different light distribution were fitted with only one component.
 
We have implemented a decision tree algorithm in order to find the best final number of structural components for each galaxy. The algorithm starts by comparing the magnitude of the galaxy obtained from the two component fit and the magnitude measured directly in the image using SExtractor. If the difference between those magnitudes is larger than 0.5 mag then, the galaxy was fitted with only one component, (this case only happened for a 2.32\% of the galaxies in our sample). This large difference between the modelled and measured magnitudes could be due to several reasons such as the presence of more than two structural components in the galaxies, or the bad  convergence of the fitted method. In the second step of the algorithm we have analysed the bulge-to-total ($B/T$) ratio given by the two component fit. Those galaxies clearly dominated by the S\'ersic components ($B/T>$0.7) were also fitted with only one component. 

\begin{figure}[h]
\centering
\includegraphics[clip,width=1.0\hsize]{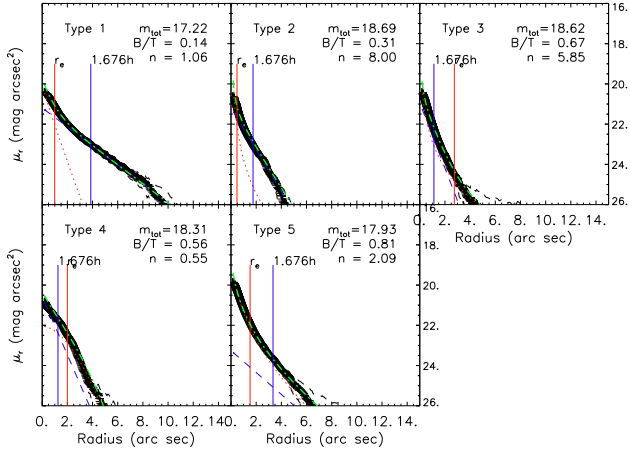}
\caption{Examples of profiles 1 to 5 following the notation of  \cite{allen06}. The black solid lines represent the galaxies surface brightness profiles and the black dashed lines show the errors limits. The red dotted lines are the S\'ersic profile fits, the blue dashed lines are the disc profile fits and the green dashed dotted lines are the sum of bulge and disc profiles fits.}
\label{fig:types}
\end{figure}

The remaining galaxies were analyzed following a similar procedure as in \cite{allen06}. We have identified five different types of fitted surface brightness profiles (see Fig \ref{fig:types}). According with the number of intersections between the S\'ersic and the exponential fitted radial profiles, we can identify those with one (Type 1, Type 2 and Type 4), two (Type 3) and zero (Type 5) intersections. Type 1 profiles were considered as 'classical' bulge plus disc galaxies. The remaining types have different features that make them to departure from a classical two component galaxy. Galaxies belonging to Type 5 show bulges dominating in the whole galaxy. Type 4 galaxies have a disc component dominating in the inner regions of the profile. Type 3 galaxies show bulge effective radius larger than the  disc effective radius (1.676$h$) and Type 2 galaxies refer to those galaxies whose S\'ersic parameter $n$ has reached the maximum value allowed in the fit.

\onecolumn

\twocolumn

Eventually, only the galaxies with Type 1 profiles were considered as two component fits. The remaining were fitted with only one component. We obtained finally that 47$\%$ of the galaxies were fitted with one S\'ersic component. 

\section{Structural parameters}

The study of correlations between the structural parameters of galaxies have been extensively investigated at low redshift in the literature (e.g. \cite{dejong96}; \cite{graham01};
    \cite{graham03}; \cite{macarthur03};\cite{mollenhoff04};\cite{aguerri04};\cite{gutierrez04};\cite{aguerri05}). Our galaxy cluster sample is located at a mean redshift of $\sim0.2$, and gives us the chance to compare the structural parameters of these galaxies with similar ones located in nearby clusters. This comparison will enable us to determine any evolution of the structural parameters of the galaxies in clusters in the last $\sim 2.5$ Gyr.
    
We have classified the galaxies taking into account the number of fitted photometrical components and their B-r colors. Three diferent galaxy types were considered: early-types (E/S0), early-spirals (Spe) and late-spirals (Spe). The results of the fit for the galaxy sample, together with their quantitative classification are shown in the Table \ref{tab:dcat} in the Appendix.

The early-type galaxies were those fitted with one S\'ersic component and located in the red sequence of the color-magnitude relations (CMR) of the clusters (within 0.2 magnitudes). Early-type spirals were those fitted with two structural components and also located near the red secuence of the CMR. Finally, late-type spirals were those objects fitted with two components and have at least 0.2 bluer B-r color than the red secuence of the cluster. This classification results than 36$\%$, 29$\%$, and 16$\%$ of the galaxies were early-type, early-spirals and late-spirals, respectively. The remaining 19$\%$ of the objects correspond to blue galaxies fitted with only one component. These objects could be a mix of different kind of objects (galaxies with more than two galactic components, blue spirals not well fitted with two components, irregular galaxies, blue ellipticals...). 

\subsection{S\'ersic Parameters}
 
In the present section we will compare the S\'ersic parameters of the galaxies in the sample  with similar galaxies located in local galaxy clusters.

\cite{kormendy77} discovered a correlation between the size and the surface brightness  $<\mu_e>$ -$r_e$ of elliptical galaxies, the so called, Kormendy Relation. Later on, \cite{binggeli84} found that this relationship was only given in elliptical galaxies brighter than $M_B\le -20$. For fainter galaxies, the tendency inverts.

In Figure \ref{fig:BulgemureALL}, we have plotted the Kormendy relation  for E/S0 (red points) and the bulges of Early Spirals (Green triangles). Notice that bulges of early-spirals and E/S0 galaxies form a continuous family of objects. The fit of this relation for both types is given by:

\begin{figure}[h]
\centering
\includegraphics[clip,width=1.0\hsize]{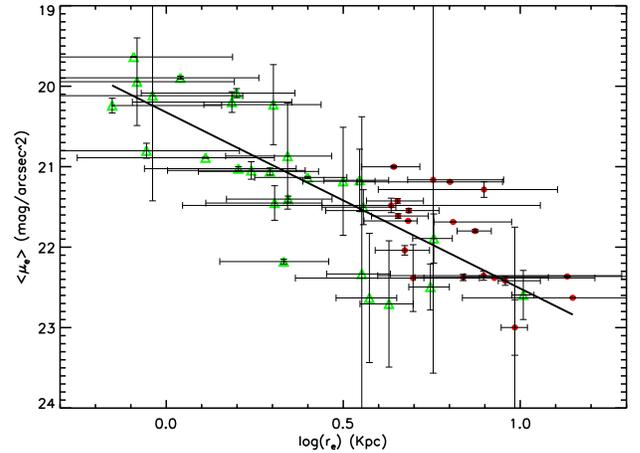}
\caption{Kormendy relation between effective ratius and mean bulge surface brightness for all the early type sample of galaxies. Red points are E/S0 galaxies, while green triangles represent the bulges of the Early Spiral galaxies. The solid line is the fit of the relation.}
\label{fig:BulgemureALL}
\end{figure}

\begin{equation}
\label{eq:kormenred}
< \mu_e>=(20.32 \pm 0.15) +(2.18 \pm 0.23)\log(r_e)
\end{equation}

This fit gives that early-type galaxies brighter than $m_{r}=20$ in clusters at z$\sim$0.2 follows the relation $L_{V}\propto r_{e}^{1.12\pm0.08}$. This relation is close to that observed for local early-type galaxies, given by $L_{B} \propto r_{e}^{1.3}$, \citep{binggeli84}.

\begin{figure}[h]
\centering
\includegraphics[clip,width=1.0\hsize]{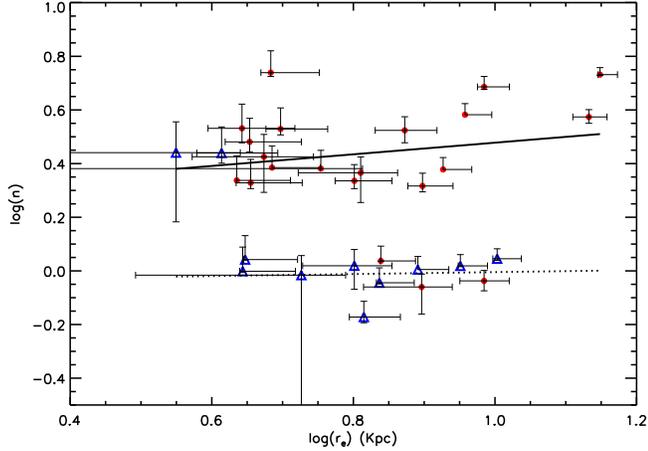}
\caption{Relation between effective ratius and shape parameter for one component galaxies. Red points refer to red-one-component galaxies and blue triangles represent blue-one-component galaxies. Solid and dotted lines are the respective fits.}
\label{fig:Bulgenre}
\end{figure}

We have also plotted in Figure \ref{fig:Bulgenre} the relation between effective-radius and shape parameters for red galaxies fitted with one component (red points) and blue galaxies fitted with one component (blue triangles). Clearly, a dichotomy exists. By taking out the obvious outliers, we have obtained the following fits.

\begin{equation}
\log n=(0.26 \pm 0.13) + (0.21 \pm 0.17 )\log(r_e) 
\end{equation}
and for the blue ones
\begin{equation}
 \log n=(-0.04 \pm 0.16) - (0.03 \pm 0.19) \log(r_e) 
\end{equation}

These fits are in agreement with recent measurements of the S\'ersic indexes in wider samples of galaxies \citep{driver06,bell08} giving support to the hypothesis that the origin and formation of these two kinds of galaxies is different. The red and blue population are located in split and parallel regions (within the errors in the slopes) in the $\log(n)-\log(r_e)$ plane. This result allows to reliably separate the early and late type galaxies by identifying the value of their S\'ersic parameter, and inversely, we can assign a particular shape to a galaxy by determining its color. As a consequence, we can conclude that the galaxies fitted with one component have a bimodal behavior. The red early-type galaxy population has an n value of $2 \leq n \leq 4$, while the blue late-type galaxy population has a shape parameter, $n \sim 1$.

We have found two blue galaxies with a S\'ersic index similar to the red galaxies in the sample. These galaxies are at 560 kpc distance from the core of A1878, a cluster which in our previous work, \cite{ascaso08} was shown to have a large blue fraction as well as an spiral dominated population. Those objects could be blue early-type galaxies results of mergers \citep{bildfell08,pipino08}, spiral galaxies with some features such as bars or dust or galaxies in process of merging.  As far as the three red galaxies with smaller S\'ersic profiles are concerned, one of them is visually classified as Spiral, so it might have been classified in one component but are real spiral galaxies, the two other ones have been classified as Ellipticals but a closer inspection of them shows that they are in clear interaction.

\begin{figure}[]
\centering
\includegraphics[clip,width=1.2\hsize]{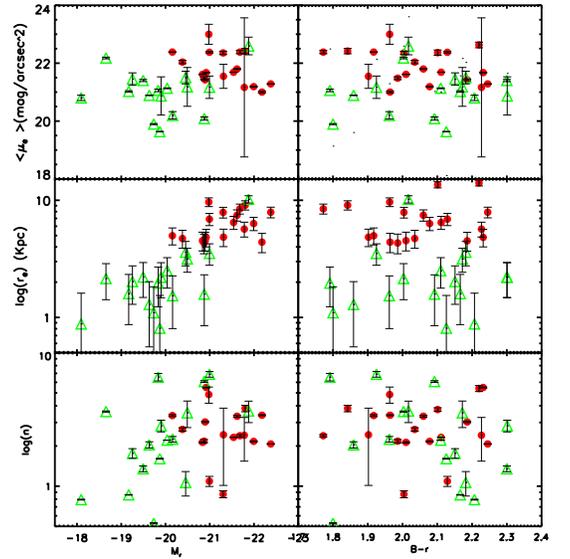}
\caption{Absolute Gunn-r Magnitude (left column) and B-r colour (right column) versus $<\mu_{e}>$, n and B/T for E/S0 (Red points) and Early Spirals Bulges (Green Triangles).}
\label{fig:pm}
\end{figure}

Furthermore, in the Figure \ref{fig:pm}, we have shown the effective surface brightness, the shape parameter and effective radius versus the absolute magnitude (left column) and color (right column) for  the E/S0 galaxies (red points) and the bulges of the Early Spiral galaxies (green triangles).  We can see that the $B-r$ colors of bulges of Early-type spirals  and E/S0 galaxies are similar (median $B-r=2.04$ and $2.06$ respectively). In general, bulges of early-type Spiral galaxies show fainter $<\mu_{e}>$ (median value is 21.05 mag arc sec$^{2}$), and smaller $r_e$ (median value is 2.02 kpc) and $n$ (median value is 2.21) S\'ersic parameters than E/S0 galaxies (median values are  21.79 mag arc sec$^{2}$, 6.46 kpc and 2.42 respectively). The plane $\log(r_e)-M_r$ shows a clear continuous relation between Early-type bulges and E/S0 galaxies as pointed before by the Kormendy relation.

Finally, we have compared the scales of our E/S0 galaxies with those from similar objects in the Coma cluster \citep{aguerri04}. Early-type Coma galaxies were selected as those galaxies with $B/T=1$ and were fitted with one component. Due to the seeing effect, we have not measured galaxies with scales smaller than 2.2 kpc. For this reason, only galaxies in Coma with re$>$2.2 kpc were considered. We obtained that the sizes of our early-type galaxies turned out to be similar to the Coma cluster galaxies (see Table \ref{tab:NOTCOMAbul}). 

As far as the shape parameter is concerned,  the range of values in the NOT sample expands all the range of values of Coma. However, we find a mean value somewhat smaller for the NOT sample than for Coma but the values agree within the errors. Therefore, it seems that the bulge sizes haven't changed substantially with respect to those in the Coma Cluster.  

\subsection{Disc Parameters}

The analysis of the structural parameters of the galaxies in clusters gives information about the role played by the environment in the evolution of galaxies in high density environments. Fast galaxy-galaxy encounters can transform galaxies from late- to early-type in short time-scales ($\sim1 Gyr$). These kind of encounters are usual in galaxy clusters, \citep{moore96}. The stars located in galactic discs have a lower binding energy than those located in the central regions of the galaxies. Thus, interactions can strip away easier stars located in the external regions of the galaxies, and truncate the discs. There are hints about this truncation in the literature. Coma cluster galaxies have shorter discs than field nearby galaxies, \citep{gutierrez04,aguerri04}.  Few facts are known about the evolution of discs at medium redshift in galaxy clusters. In the present section, we have compared the discs of the galaxies located in our medium redshift clusters with similar ones in the local environment.

\begin{figure}[]
\centering 
\includegraphics[clip,width=1.0\hsize]{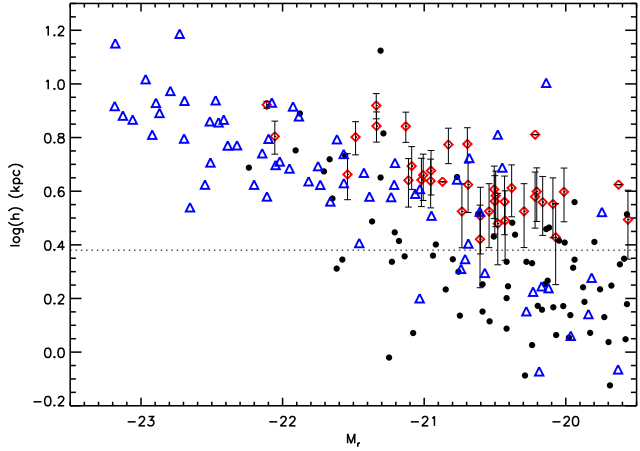}
\caption{Disc scales  versus absolute Gunn-r Magnitude for two component galaxies (Coma Cluster; \cite{aguerri04}, black points), NOT sample (red diamonds) and (field galaxies; \cite{graham01}, blue triangles). The horizontal dotted line shows the minimum disc size determined by the minimum area and surface brightness limit.}
\label{fig:freeman}
\end{figure}

We have plotted in Fig \ref{fig:freeman} the absolute magnitudes of the disks versus their scale parameters. The red diamonds concerns our medium redshift galaxy cluster sample. The blue triangles refer to a sample of field galaxies extracted from a work by \cite{graham01} in R band and the black points are the disks from the Coma cluster taken from \cite{aguerri04}. The galaxies in the latter sample have been selected as the galaxies with $B/T<1$ (two component galaxies). There are no discs in the medium redshift clusters with scales $h<2.4$ kpc. This is due to our minimum cut area (800 pixels) and to the surface brightness limit fitted corresponding to $\mu_{r} \approx 25.3 mag \, arcsec^{-2}$.  The horizontal dotted line overplotted in Fig. \ref{fig:freeman} shows this limit.

Let us note that our disc scales are as large as those of field galaxies, while those discs in Coma represent a minimum percentage. Figure \ref{fig:freeman} represents the well known Freeman law, \citep{freeman70}. The fit of this law for our disk sample is given by:

 \[ \log h=(-2.52 \pm 0.57) - (0.152 \pm 0.027) M_r\]

Regarding to a quantitative description of the disc scales,  we have found that in the central regions of our clusters at medium redshift ($R<200$ kpc) there is a population of large disc galaxies that is absent in the Coma Cluster as collected in Table \ref{tab:NOTCOMAdis}. Those results may agree with an evolution hypothesis from this range of redshift clusters to the present in the disc scales of the late type galaxy population in clusters. 

\begin{figure}[]
\centering
\includegraphics[clip,width=1.0\hsize]{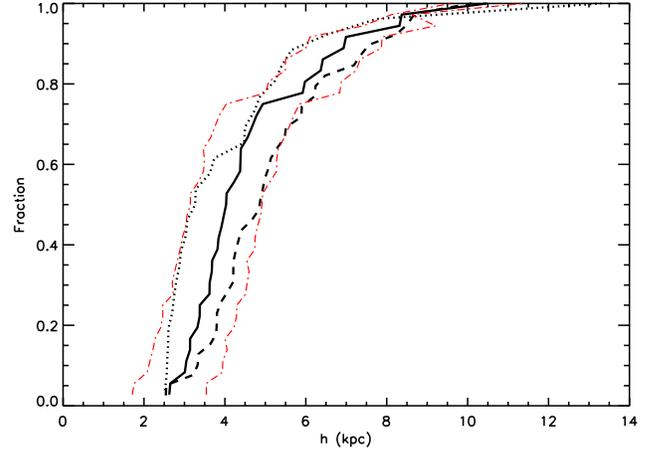}
	\caption[Cumulative function of disc scales for the NOT sample,  \cite{aguerri04} Coma Cluster and \cite{graham01} isolated  sample]{Cumulative function of disc scales for the NOT sample (solid  line), \cite{aguerri04} Coma Cluster (dotted line) and  \cite{graham01} isolated sample (dashed line). The red dotted-dashed lines show the error of the cumulative function for the NOT sample.}
\label{fig:plotsDiscFD}
\end{figure}

We have performed statistical tests to check if the disc scales in the NOT medium redshift cluster sample are significantly different to disc scales at nearby galaxies in clusters or in field. With that purpose, we have run the  Kolmogorov-Smirnov (KS) test to  the cumulative functions of the disc scales for the clusters in NOT sample, the Coma sample and the sample of isolated galaxies from \cite{graham01}. We have selected only galaxies in the same range of magnitudes. These cumulative functions are shown in Figure \ref{fig:plotsDiscFD}.

The results of the test gives that the disc scale distribution of Coma galaxies is statistically different from the disc scale distribution of both our medium redshift clusters (98.1\% significance) and the field galaxy sample (99.9\% significance). However, the disc scales of the field galaxy sample and our medium redshift galaxy sample are not statistically different (86.9\% significance).  

The distributions for Coma and NOT sample intersects at disc scales of 4 kpc.  At the view of Figure \ref{fig:plotsDiscFD}, it seems to be a larger difference between the distributions for disc sizes smaller than 4 kpc. We have performed a new Kolmogorov-Smirnov test for galaxies with discs larger than 4 kpc in all samples. The results show that none of the samples can be ensured to be statistically different with 54.95\% significance for the Coma Cluster and NOT sample, 28.95\% for the Coma Cluster and field galaxy sample and 28.37\% for the NOT sample and field galaxy sample. 

If we consider in our analysis the galaxies classified visually as Spirals in \cite{ascaso08}, we find similar results. The whole disc scale distributions for Coma and NOT clusters are statistically different with a (98.98\% significance), while for NOT clusters and the field galaxy sample are not statistically different (19.41\% significance). In addition, by considering only the galaxies with disc scales larger than 4 kpc in this subsample, the disc scales distributions for NOT and Coma galaxy clusters and NOT and field galaxies are not statistically different with 66.55\% and 28.95\% respectively. This result reinforces the difference in the smaller disc scales distributions in local and medium redshift clusters from a morphological point of view.

As a conclusion, Coma shows an excess of galaxies with discs scales smaller than 4 kpc with respect to the field galaxy and NOT sample.  Field galaxies with large disc could have entered in the cluster and suffered interactions with the environment that have made the discs get shorter \citep{aguerri05,aguerri09,kormendy08} in the last 2.5 Gyr. This conclusion supports an scenario in which the environment plays a crucial role in the galaxy evolution.

\section{Discussion and Conclusions}

In this paper, we have analyzed the structural properties of a sample of galaxies placed in a five clusters between 0.18 $\leq z \leq$ 0.25. The structural parameters derived from these quantitative classification have shown us that the galaxies fitted into one component have a nearly univocally correspondence with the shape parameter. Red galaxies fitted with one component have a shape parameter between 2 and 4, while for the blue galaxies fitted with one component, the shape parameter we obtain is around 1. This bi-modality has been already observed in previous works \citep{driver06,bell08} showing a clear difference in the formation and evolution of this two types of galaxy population.

Likewise, we have looked into the relation of the structural parameters. We have looked into the Kormendy Relation and we have found that the E/S0 and the bulges of early Spirals seem to be a continuous set of parameters. Similar results are obtained  with magnitude and color in Figure \ref{fig:pm}. The results show that  $r_e$ and $n$ follow a correlation with the magnitude taking out the obvious outlier for the early types and for the bulges of early spirals, suggesting a similar nature of this objects.  
 
On the other hand, we have compared the bulge scales of the galaxies in the sample with scale of the bulges of galaxies in the Coma cluster \citep{aguerri04}, obtaining the same range of values. As far as the disc scales of the galaxies in our medium redshift sample are concerned, we have compared them with the disc scales in a sample of galaxies in Coma studied by \cite{aguerri04} and the disc scales of a sample of local field galaxies studied by \cite{graham01}. We have considered the whole galaxy sample classified in two components following the procedure explained in 3.3 and also, to make sure that we were comparing spiral visually classified galaxies, we have taken the whole galaxy sample classified as Spiral galaxy in \cite{ascaso08}. We have found in both cases that the disc scales of the galaxies in our sample are slightly larger than those in Coma cluster and statistically different. Moreover, the distribution of disc scales for the NOT sample is not statistically different to the disc scales for the local field galaxies  This result indicates an evolution in the disc scales of the galaxies in clusters at medium redshift from local clusters. 


Interestingly, we have seen that the main difference of the distribution is concentrated on the small disc galaxies ($<$ 4 $kpc$). Thus, this fact shows a different behavior between small and large disc galaxies, indicating a different nature of these kind of objects. There is a large population of galaxies with $h<4$ kpc in the Coma cluster not observed in the field or in the medium redshift clusters. 

One of the main concerns to the present work could be that the compared galaxy clusters at z$\sim$0 (the Coma cluster) and at z$\sim$0.25 (our NOT sample) are not statistical complete samples. It can be argued that the Coma cluster is not a representative comparison due to its high mass and degree of evolution. Thus, the different disc scales founded between the Coma cluster and those at medium redshift could be due to the fact that we are comparing clusters with very different properties. We have tested this assumption by splitting our clusters sample into two groups according with their galaxy richness: poor clusters (Richness Class (RC) $<$2) and rich clusters (RC $\ge$ 2). The Coma cluster has a richness class similar to our rich clusters sample at z$\sim$0.25 (RC=2). We have found that, for both samples, their disc scale length distributions are different from Coma and similar to the field one. This is the same tendency that is observed for the overall distribution. Therefore,  the result does not depend on the richness of the clusters. 

More work needs to be performed in the future to span both cluster samples and determine the variance in the disc scale distribution in nearby and medium redshift clusters. Nevertheless, this result indicates that there was a strong evolution in the external parts of the galaxies in clusters during the last 2.5 Gyr or that Coma is in some way anomalous. Mechanisms like galaxy harassment with small timescales ($\sim$ 1 Gyr) can explain this evolution.


\begin{acknowledgements}

\end{acknowledgements}
We acknowledge the anonymous referee for improving this paper. Bego\~na Ascaso is partially supported by NASA grant NNG05GD32G. JALA and RSJ were partially supported by the Ministerio de Ciencia e innovaci\'on by the grants AYA2007-67965-C03-01 and CSD2006-00070

 \begin{table*}[!h]
      \caption[]{The sample of Clusters}
      \[
         \begin{array}{llllllllccclll}
            \hline\noalign{\smallskip}
\multicolumn{1}{c}{\rm Name}&
\multicolumn{3}{c}{\alpha (2000)}&
\multicolumn{3}{c}{\delta (2000)}&
\multicolumn{1}{c}{z}&
\multicolumn{1}{c}{\# frames}&
\multicolumn{1}{c}{\rm Area (Mpc}$^2$)&
\multicolumn{1}{c}{\rm seeing \, ('')}\\
\hline\noalign{\smallskip}
{\rm A~2658} & 23 & 44 & 49 & -12 & 17 & 39 & 0.185 & 1 & 0.3055   & 0.70 \\
{\rm A~1643} & 12 & 55 & 54 & +44 & 05 & 12 & 0.198 & 2 & 0.6810  & 0.55 \\
{\rm A~1878} & 14 & 12 & 52 & +29 & 14 & 28 & 0.220 & 2 &  0.7894 & 0.70  \\
{\rm A~2111} & 15 & 39 & 40 & +34 & 25 & 27 & 0.229 & 2 & 0.8030  & 0.70 \\
{\rm A~1952} & 14 & 41 & 03 & +28 & 37 & 00 & 0.248 & 2 & 0.7989  & 0.55-0.80\\
\hline
         \end{array}
      \]
\label{tab:all}
   \end{table*}

\begin{table*}[]
      \caption{Bulge parameters for Coma and NOT sample}
      \[
         \begin{array}{lcccccc}
            \hline\noalign{\smallskip}
\multicolumn{1}{c}{\rm Name}&
\multicolumn{1}{c}{\rm <r_e>}&
\multicolumn{1}{c}{\rm \sigma (r_e)}&
\multicolumn{1}{c}{\rm <n>}&
\multicolumn{1}{c}{\rm \sigma (n)}&
\multicolumn{1}{c}{\rm < Dist (Kpc)>}&
\multicolumn{1}{c}{\rm \sigma (Dist (Kpc))}\\
\hline\noalign{\smallskip}
{\rm NOT } & 6.58 & 2.38 & 2.24 & 1.35 & 349.72& 257.053\\ 
{\rm Coma} & 8.06 & 16.34 & 3.49 & 1.60 & 410.684 & 243.904 \\ 
\hline
         \end{array}
      \]
\label{tab:NOTCOMAbul}
   \end{table*}

\begin{table*}[]
      \caption{Disc parameters for Coma and NOT sample}
      \[
         \begin{array}{lcccccc}
            \hline\noalign{\smallskip}
\multicolumn{1}{c}{\rm Name}&
\multicolumn{1}{c}{\rm <h>}&
\multicolumn{1}{c}{\rm \sigma (h)}&
\multicolumn{1}{c}{\rm < Dist (Kpc)>}&
\multicolumn{1}{c}{\rm \sigma (Dist (Kpc))}\\
\hline\noalign{\smallskip}
{\rm NOT } & 4.738 & 1.941 &  272.16 & 202.10\\ 
{\rm Coma} & 4.367 & 2.899 &  505.008 & 259.386 \\ 
\hline
         \end{array}
      \]
\label{tab:NOTCOMAdis}
   \end{table*}

\onecolumn
\begin{appendix}
\section{Results of the 2-dimensional surface brightness fit for the galaxies in NOT sample}
\begin{landscape}
\tiny 
\begin{longtable}{lcccccccccccccccccc}
            \hline
\multicolumn{1}{c}{\rm Name}&
\multicolumn{1}{c}{$\alpha$ (J2000)}&
\multicolumn{1}{c}{$\delta$ (J2000)}&
\multicolumn{1}{c}{z}&
\multicolumn{1}{c}{$B-r$}&
\multicolumn{1}{c}{M$_{r}$}&
\multicolumn{1}{c}{$\mu_e$}&
\multicolumn{1}{c}{$r_e$}&
\multicolumn{1}{c}{$e_b$}&
\multicolumn{1}{c}{$\mu_0$}&
\multicolumn{1}{c}{$h$}&
\multicolumn{1}{c}{$e_d$}&
\multicolumn{1}{c}{$n$}&
\multicolumn{1}{c}{$PA_b$}&
\multicolumn{1}{c}{$PA_d$}&
\multicolumn{1}{c}{$B/T$}&
\multicolumn{1}{c}{$\chi^2$}&
\multicolumn{1}{c}{$Col$}&
\multicolumn{1}{c}{T}\\
\multicolumn{1}{c}{}&
\multicolumn{1}{c}{h:m:s}&
\multicolumn{1}{c}{d:m:s}&
\multicolumn{1}{c}{}&
\multicolumn{1}{c}{}&
\multicolumn{1}{c}{}&
\multicolumn{1}{c}{mag/arc sec$^2$}&
\multicolumn{1}{c}{arc sec}&
\multicolumn{1}{c}{}&
\multicolumn{1}{c}{mag/arc sec$^2$}&
\multicolumn{1}{c}{arc sec}&
\multicolumn{1}{c}{}&
\multicolumn{1}{c}{}&
\multicolumn{1}{c}{}&
\multicolumn{1}{c}{}&
\multicolumn{1}{c}{}&
\multicolumn{1}{c}{}&
\multicolumn{1}{c}{}&
\multicolumn{1}{c}{}\\
\hline\noalign{\smallskip}
{\rm A~1643} &        12:55:55.18  &	  44:03:47.50 & 	       &   2.013 &     -20.67 &    24.06	&     1.15597  &       0.739 &   21.14   &    1.10000	 &    0.586 &	    8.000    &  169.217 &      11.828 &      0.324 &	  2.308 &	R &    S0   \\    
{\rm A~1643} &        12:55:47.93  &	  44:04:01.20 & 	       &   2.035 &     -20.36 &    22.91	&     1.45499  &       0.756 &    0.00   &    0.00000	 &    0.000 &	    2.663    &    0.087 &	0.000 &      1.000 &	  1.089 &	R &	E   \\    
{\rm A~1643} &        12:55:50.96  &	  44:04:31.00 & 	       &   1.930 &     -21.15 &    21.13	&    0.217008  &       0.449 &   21.20   &    1.46502	 &    0.775 &	    8.000    &  137.578 &      41.868 &      0.065 &	  2.168 &	R &	E   \\    
{\rm A~1643} &        12:55:55.35  &	  44:04:34.40 & 	       &   1.963 &     -20.69 &    24.49	&     2.97792  &       1.000 &    0.00   &    0.00000	 &    0.000 &	    4.854    &   88.788 &	0.000 &      1.000 &	  3.292 &	R &	E   \\    
{\rm A~1643} &        12:55:52.70  &	  44:04:44.50 & 	       &   1.818 &     -20.42 &    23.10	&     1.31296  &       0.285 &   21.34   &   0.962016	 &    0.564 &	    8.000    &   57.503 &      50.222 &      0.485 &	  1.915 &	R &    S0   \\    
{\rm A~1643} &        12:55:59.29  &	  44:04:57.10 & 	       &   2.011 &     -20.02 &    23.02	&     2.97704  &       0.192 &    0.00   &    0.00000	 &    0.000 &	    0.918    &  126.466 &	0.000 &      1.000 &	  2.328 &	R &	S   \\    
{\rm A~1643} &        12:55:54.00  &	  44:05:12.40 & 	       &   2.060 &     -21.61 &    22.69	&     2.29997  &       0.691 &    0.00   &    0.00000	 &    0.000 &	    3.341    &    1.579 &	0.000 &      1.000 &	  1.515 &	R &    S0   \\    
{\rm A~1643} &        12:55:45.24  &	  44:06:34.00 & 	       &   2.301 &     -21.74 &    21.64	&    0.679008  &       0.670 &   22.06   &    2.56203	 &    0.793 &	    2.835    &   12.302 &      46.929 &      0.210 &	  1.628 &	R &	S   \\    
{\rm A~1643} &        12:55:33.82  &	  44:07:12.50 & 	       &   2.186 &     -20.93 &    22.53	&     1.38899  &       0.879 &    0.00   &    0.00000	 &    0.000 &	    3.027    &  108.131 &	0.000 &      1.000 &	  2.479 &	R &	E   \\    
{\rm A~1643} &        12:55:36.40  &	  44:07:53.40 & 	       &   1.017 &     -20.71 &    22.56	&     2.75598  &       0.318 &    0.00   &    0.00000	 &    0.000 &	    1.044    &   67.353 &	0.000 &      1.000 &	  7.040 &	B &	I   \\    
{\rm A~1643} &        12:55:36.57  &	  44:08:30.40 & 	       &   1.917 &     -20.27 &    23.47	&     1.53595  &       0.819 &    0.00   &    0.00000	 &    0.000 &	    3.380    &   12.872 &	0.000 &      1.000 &	  1.532 &	R &	E   \\    
{\rm A~1878} &        14:12:47.82  &	  29:13:53.40 & 	       &   2.111 &     -21.69 &    22.06	&     1.83603  &       0.508 &    0.00   &    0.00000	 &    0.000 &	    2.321    &    0.688 &	0.000 &      1.000 &	  3.025 &	R &	S   \\    
{\rm A~1878} &        14:12:49.47  &	  29:14:09.90 & 	       &   2.151 &     -21.57 &    21.43	&    0.574992  &       0.404 &   20.89   &    1.24203	 &    0.807 &	    1.763    &   70.882 &      20.411 &      0.145 &	  1.341 &	R &	S   \\    
{\rm A~1878} &        14:12:52.50  &	  29:14:11.40 & 	       &   1.701 &     -20.94 &    22.03	&     1.25101  &       0.725 &    0.00   &    0.00000	 &    0.000 &	    0.997    &   49.960 &	0.000 &      1.000 &	  2.567 &	B &	S   \\    
{\rm A~1878} &        14:12:54.15  &	  29:14:19.30 & 	       &   2.207 &     -20.80 &    20.89	&    0.249920  &       0.633 &   20.82   &   0.880000	 &    0.807 &	    0.793    &   22.695 &      34.053 &      0.099 &	  1.132 &	R &	E   \\    
{\rm A~1878} &        14:12:52.18  &	  29:14:28.40 &      0.2220    &   2.300 &     -22.36 &    21.35	&    0.627968  &       0.441 &   20.71   &    1.80998	 &    0.770 &	    1.350    &   45.974 &	1.528 &      0.081 &	  2.285 &	R &	E   \\    
{\rm A~1878} &        14:12:46.85  &	  29:14:26.40 & 	       &   1.655 &     -21.02 &    22.47	&     1.85504  &       0.554 &    0.00   &    0.00000	 &    0.000 &	    0.673    &   82.918 &	0.000 &      1.000 &	  2.476 &	B &	I   \\    
{\rm A~1878} &        14:12:54.72  &	  29:14:31.90 & 	       &   2.172 &     -21.42 &    22.16	&    0.897952  &       0.726 &   21.24   &    1.14893	 &    0.745 &	    3.544    &   34.265 &      62.113 &      0.481 &	  1.199 &	R &	E   \\    
{\rm A~1878} &        14:12:50.98  &	  29:14:42.30 & 	       &   2.003 &     -20.84 &    22.59	&    0.610016  &       0.425 &   21.51   &    1.16301	 &    0.835 &	    3.623    &  167.017 &      31.290 &      0.160 &	  1.736 &	R &	I   \\    
{\rm A~1878} &        14:12:49.12  &	  29:14:42.50 & 	       &   2.135 &     -21.38 &    24.25	&     1.58400  &       1.000 &   20.57   &    1.19698	 &    0.438 &	    8.000    &   63.119 &      38.637 &      0.420 &	  2.355 &	R &	S   \\    
{\rm A~1878} &        14:12:50.96  &	  29:14:56.60 & 	       &   2.166 &     -21.29 &    21.56	&    0.454080  &       0.918 &   20.99   &    1.68802	 &    0.369 &	    0.861    &  147.586 &	4.530 &      0.169 &	  1.703 &	R &	S   \\    
{\rm A~1878} &        14:12:46.58  &	  29:14:59.10 & 	       &   2.280 &     -20.94 &    23.28	&     1.01394  &       0.476 &   20.57   &   0.951984	 &    0.481 &	    8.000    &  160.326 &      48.481 &      0.332 &	  1.774 &	R &    S0   \\    
{\rm A~1878} &        14:12:52.43  &	  29:15:48.70 & 	       &   2.212 &     -21.00 &    20.88	&    0.259952  &       0.402 &   20.32   &   0.915024	 &    0.550 &	    8.000    &    8.221 &      59.070 &      0.159 &	  1.789 &	R &    S0   \\    
{\rm A~1878} &        14:13:00.54  &	  29:13:56.90 & 	       &   1.771 &     -21.15 &    23.67	&     1.46397  &       0.812 &   21.47   &    1.12499	 &    0.610 &	    5.988    &  143.886 &     133.731 &      0.582 &	  1.324 &	B &    S0   \\    
{\rm A~1878} &        14:12:57.80  &	  29:12:01.60 & 	       &   1.508 &     -20.47 &    22.77	&     1.26104  &       0.911 &    0.00   &    0.00000	 &    0.000 &	    1.102    &  114.178 &	0.000 &      1.000 &	  3.744 &	B &    S0   \\    
{\rm A~1878} &        14:12:59.05  &	  29:12:14.40 & 	       &   1.611 &     -20.60 &    22.26	&     1.00707  &       0.697 &    0.00   &    0.00000	 &    0.000 &	    2.759    &  177.224 &	0.000 &      1.000 &	  2.358 &	B &	E   \\    
{\rm A~1878} &        14:12:59.84  &	  29:12:19.50 & 	       &   1.750 &     -20.45 &    21.68	&     1.12006  &       0.065 &   23.00   &    2.97106	 &    0.214 &	    2.400    &   53.622 &      52.732 &      0.305 &	  4.800 &	B &	S   \\    
{\rm A~1878} &        14:13:00.58  &	  29:12:22.90 & 	       &   1.368 &     -20.30 &    22.99	&     1.51395  &       0.661 &    0.00   &    0.00000	 &    0.000 &	    0.964    &  145.345 &	0.000 &      1.000 &	  2.068 &	B &	S   \\    
{\rm A~1878} &        14:13:05.59  &	  29:12:54.20 & 	       &   1.675 &     -20.53 &    21.89       &     0.815056 &       0.428 &    21.50  &     1.02907	&    0.864 &	   1.152    &  111.972 &      61.071 &      0.318 &	 1.645 &       B &     E   \\	 
{\rm A~1878} &        14:13:02.81  &	  29:12:55.70 & 	       &   1.869 &     -19.44 &    22.01       &      1.16794 &       0.393 &     0.00  &     0.00000	&    0.000 &	   2.751    &  135.302 &       0.000 &      1.000 &	 3.636 &       B &     S   \\	 
{\rm A~1952} &        14:41:07.84  &	  28:38:29.40 & 	       &   2.102 &     -22.05 &    23.20       &      1.47893 &       0.662 &    20.61  &     1.64402	&    0.774 &	   8.000    &	16.433 &      53.677 &      0.402 &	 1.129 &       R &     E   \\	 
{\rm A~1952} &        14:41:01.92  &	  28:37:14.50 & 	       &   2.126 &     -20.76 &    18.11       &     0.209968 &       0.105 &    20.24  &    0.696080	&    0.838 &	   1.606    &  104.714 &      74.489 &      0.285 &	 0.955 &       R &     E   \\	 
{\rm A~1952} &        14:41:02.66  &	  28:37:10.00 & 	       &   2.220 &     -22.11 &    23.84       &      3.65200 &       0.734 &     0.00  &     0.00000	&    0.000 &	   5.391    &	29.119 &       0.000 &      1.000 &	 1.835 &       R &    S0   \\	 
{\rm A~1952} &        14:41:03.13  &	  28:37:10.10 & 	       &   1.964 &     -21.41 &    22.12       &      1.13995 &       0.843 &     0.00  &     0.00000	&    0.000 &	   3.400    &  108.770 &       0.000 &      1.000 &	 1.795 &       R &     E   \\	 
{\rm A~1952} &        14:41:01.19  &	  28:37:00.50 & 	       &   2.061 &     -21.20 &    21.51       &     0.520960 &       0.649 &    20.07  &    0.947936	&    0.575 &	   8.000    &	21.700 &      61.873 &      0.489 &	 0.924 &       R &     E   \\	 
{\rm A~1952} &        14:41:01.32  &	  28:37:43.20 & 	       &   2.092 &     -21.57 &    21.02       &     0.410080 &       0.536 &    20.63  &     1.28198	&    0.902 &	   6.079    &  156.086 &      73.545 &      0.282 &	 0.780 &       R &     E   \\	 
{\rm A~1952} &        14:41:13.59  &	  28:37:29.60 & 	       &   1.786 &     -22.13 &    22.45	&     1.51800  &       0.742 &   19.59   &    1.12798	 &    0.395 &	    4.693    &  157.794 &     140.539 &      0.666 &	  1.639 &	  B &	 S0   \\    
{\rm A~1952} &        14:41:14.94  &	  28:37:42.60 & 	       &   1.877 &     -21.80 &    20.44	&    0.677072  &       0.590 &   20.77   &    1.80400	 &    0.536 &	    1.200    &   68.440 &      79.062 &      0.475 &	  1.825 &	B &    S0   \\    
{\rm A~1952} &        14:41:07.59  &	  28:35:35.00 & 	       &   1.733 &     -21.58 &    21.86       &      2.61395 &       0.211 &     0.00  &     0.00000	&    0.000 &	   1.110    &	98.621 &       0.000 &      1.000 &	13.768 &       B &     S   \\	 
{\rm A~1952} &        14:41:08.19  &	  28:35:44.50 & 	       &   1.567 &     -21.61 &    22.46       &      1.05406 &       0.404 &    20.49  &     1.18694	&    0.860 &	   6.009    &	28.048 &      28.236 &      0.356 &	 3.480 &       B &    S0   \\	 
{\rm A~1952} &        14:41:07.03  &	  28:36:39.20 & 	       &   0.504 &     -22.10 &    16.91       &     0.186912 &       0.308 &    19.86  &    0.781968	&    0.681 &	   2.343    &  168.989 &      83.230 &      0.696 &	 7.501 &       B &    S0   \\	 
{\rm A~1952} &        14:41:03.11  &	  28:36:46.60 & 	       &   1.859 &     -20.74 &    19.60       &     0.334928 &       0.117 &    20.41  &    0.994928	&    0.704 &	   2.041    &	96.391 &      88.057 &      0.178 &	 1.366 &       R &    S0   \\	 
{\rm A~1952} &        14:41:03.57  &	  28:37:00.30 & 	       &   2.101 &     -22.61 &    23.65       &      3.52194 &       0.937 &     0.00  &     0.00000	&    0.000 &	   3.750    &  105.293 &       0.000 &      1.000 &	 1.711 &       R &     E   \\	 
{\rm A~1952} &        14:41:08.25  &	  28:37:13.80 & 	       &   1.676 &     -21.85 &    20.84       &     0.687984 &       0.604 &    20.38  &     1.13802	&    0.803 &	   2.041    &	49.162 &     106.764 &      0.482 &	 3.233 &       B &     S   \\	 
{\rm A~2111} &        15:39:37.64  &	  34:27:03.80 &      0.2295    &   2.110 &     -21.26 &    21.00	&    0.690976  &       0.326 &   20.99   &   0.922944	 &    0.810 &	    2.211    &   20.762 &      24.650 &      0.375 &	  1.441 &	R &    S0   \\    
{\rm A~2111} &        15:39:40.49  &	  34:25:27.30 &      0.2282    &   2.182 &     -22.67 &    22.02	&    0.992992  &       0.830 &   21.21   &    2.30402	 &    0.672 &	    1.065    &  174.278 &      41.050 &      0.172 &	  1.530 &	R &	E   \\    
{\rm A~2111} &        15:39:39.20  &	  34:25:11.50 & 	       &   2.129 &     -21.13 &    22.85	&     1.90098  &       0.787 &    0.00   &    0.00000	 &    0.000 &	    1.089    &   83.748 &	0.000 &      1.000 &	  7.623 &	R &	E   \\    
{\rm A~2111} &        15:39:39.39  &	  34:25:13.40 &      0.2211    &   2.004 &     -21.34 &    22.78	&     2.17096  &       0.831 &    0.00   &    0.00000	 &    0.000 &	    0.871    &   20.000 &	0.000 &      1.000 &	  5.658 &	R &	E   \\    
{\rm A~2111} &        15:39:36.79  &	  34:25:39.10 &      0.2312    &   2.232 &     -20.90 &    22.41	&     1.32898  &       0.469 &    0.00   &    0.00000	 &    0.000 &	    5.484    &    4.038 &	0.000 &      1.000 &	  4.855 &	R &    S0   \\    
{\rm A~2111} &        15:39:34.26  &	  34:26:12.50 &      0.2289    &   2.227 &     -21.97 &    22.12	&     1.56394  &       0.859 &    0.00   &    0.00000	 &    0.000 &	    2.406    &  134.942 &	0.000 &      1.000 &	  3.430 &	R &    S0   \\    
{\rm A~2111} &        15:39:38.70  &	  34:26:38.80 &      0.2246    &   1.477 &     -20.85 &    22.20	&     1.89200  &       0.311 &    0.00   &    0.00000	 &    0.000 &	    0.903    &   20.000 &	0.000 &      1.000 &	  2.158 &	B &	S   \\    
{\rm A~2111} &        15:39:41.34  &	  34:24:34.30 &      0.2294    &   2.111 &     -20.97 &    22.11	&    0.479072  &       0.528 &   21.21   &    1.04597	 &    0.569 &	    8.000    &  178.202 &      62.997 &      0.296 &	  1.561 &	R &	S   \\    
{\rm A~2111} &        15:39:41.81  &	  34:24:42.70 &      0.2292    &   2.245 &     -22.61 &    22.25	&     2.17694  &       0.929 &    0.00   &    0.00000	 &    0.000 &	    2.072    &  142.212 &	0.000 &      1.000 &	  1.469 &	R &	E   \\    
{\rm A~2111} &        15:39:47.09  &	  34:27:37.90 &      0.2368    &   1.666 &     -21.25 &    21.49	&    0.652960  &       0.268 &   20.70   &   0.922064	 &    0.735 &	    2.042    &  153.304 &     143.414 &      0.184 &	  1.712 &	B &    S0   \\    
{\rm A~2111} &        15:39:52.99  &	  34:27:48.60 &      0.2297    &   2.010 &     -20.98 &    21.85	&     1.24502  &       0.468 &    0.00   &    0.00000	 &    0.000 &	    2.130    &   81.730 &	0.000 &      1.000 &	  2.925 &	R &    S0   \\    
{\rm A~2111} &        15:39:42.02  &	  34:26:30.30 &      0.2258    &   2.078 &     -22.09 &    22.23	&     1.74398  &       0.975 &    0.00   &    0.00000	 &    0.000 &	    2.166    &   10.666 &	0.000 &      1.000 &	  0.869 &	R &	E   \\    
{\rm A~2111} &        15:39:49.35  &	  34:26:41.50 &      0.2299    &   1.548 &     -21.54 &    22.38	&     1.74398  &       0.734 &    0.00   &    0.00000	 &    0.000 &	    1.045    &   14.595 &	0.000 &      1.000 &	  4.741 &	B &	S   \\    
{\rm A~2111} &        15:39:45.75  &	  34:26:57.40 &      0.2292    &   1.965 &     -21.07 &    18.86	&    0.227920  &       0.255 &   20.57   &   0.726000	 &    0.936 &	    0.500    &   49.420 &      54.609 &      0.156 &	  1.207 &	R &	E   \\    
{\rm A~2111} &        15:39:52.15  &	  34:27:12.20 & 	       &   1.050 &     -21.13 &    23.31	&     2.14403  &       0.728 &    0.00   &    0.00000	 &    0.000 &	    1.012    &   35.048 &	0.000 &      1.000 &	  1.646 &	B &	S   \\    
{\rm A~2111} &        15:39:47.34  &	  34:25:10.20 &      0.2309    &   1.986 &     -21.07 &    22.47	&     1.18994  &       0.924 &    0.00   &    0.00000	 &    0.000 &	    2.174    &  126.333 &	0.000 &      1.000 &	  1.065 &	R &	E   \\    
{\rm A~2658} &        23:44:50.35  &	 -12:18:25.50 & 	       &   1.800 &     -21.89 &    19.13	&    0.356928  &       0.335 &   20.00   &    1.49706	 &    0.656 &	    0.527    &   97.394 &      55.478 &      0.115 &	  8.090 &	R &	S   \\    
{\rm A~2658} &        23:44:46.97  &	 -12:18:10.40 & 	       &   1.962 &     -20.94 &    21.29	&    0.498960  &       1.000 &   20.81   &    1.29202	 &    0.545 &	    2.246    &  134.927 &     160.502 &      0.398 &	  4.597 &	R &    S0   \\    
{\rm A~2658} &        23:44:54.27  &	 -12:17:59.30 & 	       &   1.925 &     -21.42 &    22.13	&     1.14594  &       0.520 &   20.71   &    1.18395	 &    0.727 &	    6.887    &   21.717 &      75.161 &      0.536 &	  6.948 &	R &	E   \\       
{\rm A~2658} &        23:44:47.44  &	 -12:17:47.40 & 	       &   1.902 &     -20.92 &    22.51	&     1.57696  &       0.862 &    0.00   &    0.00000	 &    0.000 &	    2.427    &   69.348 &	0.000 &      1.000 &	  4.363 &	R &	E   \\       
{\rm A~2658} &        23:44:50.26  &	 -12:17:20.90 & 	       &   1.842 &     -21.07 &    23.36	&     2.95398  &       0.679 &    0.00   &    0.00000	 &    0.000 &	    3.821    &   31.148 &	0.000 &      1.000 &	  3.693 &	R &    S0   \\       
{\rm A~2658} &        23:44:49.84  &	 -12:17:26.70 & 	       &   1.791 &     -21.32 &    22.49	&    0.639056  &       0.815 &   21.60   &    1.94304	 &    0.784 &	    6.561    &   41.131 &      29.707 &      0.238 &	  5.542 &	R &	E   \\       
{\rm A~2658} &        23:44:49.80  &	 -12:17:39.50 & 	       &   2.017 &     -22.39 &    23.60	&     3.32094  &       0.731 &   21.34   &    2.26794	 &    0.814 &	    3.665    &   10.701 &      61.252 &      0.531 &	  3.578 &	R &	E   \\       
{\rm A~2658} &        23:44:56.18  &	 -12:17:07.50 & 	       &   1.772 &     -21.14 &    23.14	&     2.75106  &       0.719 &    0.00   &    0.00000	 &    0.000 &	    2.388    &   11.523 &	0.000 &      1.000 &	 11.770 &	R &	S   \\       
\hline
\label{tab:dcat}
\end{longtable}
\begin{minipage}{230mm}
NOTE. Col. (1): Galaxy Cluster; Col. (2): Right ascension (J2000); Col. (3): Declination (J2000);
Col. (4): Redshift; Col. (5): B-r color; Col. (6): Gunn-r Absolute magnitude; Col. (7): Effective surface brightness of the bulge; Col. (8): Effective radius of the bulge;  Col. (9): Ellipticity of the bulge; Col. (10): Central surface brightness of the disk; Col. (11): Scale length of the disk; Col. (12): Ellipticity of the disk; Col. (13): Shape parameter of the bulge; Col. (14): Position angle of the bulge; Col. (15): Position angle of the disk; Col. (16): bulge-to-total luminosity ratio; Col. (17): $\chi^2$ of the fit; Col. (18): Assigned color. Red (R) galaxies are located within 0.2 magnitudes from the CMR of the host cluster and Blue (B) galaxies are bluer than 0.2 mag the CMR of the cluster;  Col. (19): Visual morphological type  extracted from \cite{ascaso08}. \end{minipage}
\end{landscape}

\end{appendix}

\end{document}